**The Architecture of AI Transformation:**

**Four Strategic Patterns and an Emerging Frontier**

Diana A. Wolfe[1], Alice Choe[2], Fergus Kidd[3]

[1] Doctoral Candidate, Industrial Organizational Psychology, Seattle Pacific University, 3307 3rd Ave. W, Seattle, WA

[2] Doctoral Candidate, Organizational Behavior and Human Resource Management, Rotman School of Management, University of Toronto, 105 St. George Street, Toronto, Ontario, Canada

[3] Head of Engineering, Avanade, 30 Cannon Street, London, UK

∗To whom correspondence should be addressed; E-mail: diana.wolfe@avanade.com

## Abstract

Despite extensive investment in artificial intelligence, 95% of enterprises report no measurable profit impact from AI deployments (MIT, 2025).  In this theoretical paper, we argue that this gap reflects paradigmatic lock-in that channels AI into incremental optimization rather than structural transformation. Using a cross-case analysis, we propose a 2×2 framework that reconceptualizes AI strategy along two independent dimensions: the degree of transformation achieved (incremental to transformational) and the treatment of human contribution (reduced to amplified). The framework surfaces four patterns now dominant in practice: individual augmentation, process automation, workforce substitution, and a less deployed frontier of collaborative intelligence. Evidence shows that the first three dimensions reinforce legacy work models and yield localized gains without durable value capture. Realizing collaborative intelligence requires three mechanisms: complementarity (pairing distinct human and machine strengths), co-evolution (mutual adaptation through interaction), and boundary-setting (human determination of ethical and strategic parameters). Complementarity and boundary-setting are observable in regulated and high-stakes domains; co-evolution is largely absent, which helps explain limited system-level impact. Our findings in a case study analysis illustrated that advancing toward collaborative intelligence requires material restructuring of roles, governance, and data architecture rather than additional tools. The framework reframes AI transformation as an organizational design challenge: moving from optimizing the division of labor between humans and machines to architecting their convergence, with implications for operating models, workforce development, and the future of work.

## Paradigmatic Lock-in: Why AI Transformation Remains Elusive

Artificial intelligence (AI) refers to computational systems capable of tasks that once required human judgment, including perception, prediction, and language generation. In practice, AI is already embedded across industries: automating customer interactions, accelerating software development, optimizing logistics, and supporting complex decision analysis. These applications demonstrate efficiency gains but rarely extend to deeper organizational redesign. AI transformation describes that next step. It is the systematic reconfiguration of work, processes, and governance so that human and machine intelligence operate together as part of an integrated system. Transformation is distinct from incremental adoption. It does not mean simply adding AI tools into existing workflows. It means restructuring roles, decision rights, and data architectures in ways that allow new forms of value creation.

While automation and augmentation have dominated both theory and practice, they frame AI's role as a binary choice, either replacing human labor or enhancing it, without offering a pathway for deeper organizational change (Raisch & Krakowski, 2021). In parallel, transformation remains an abstract aspiration rather than a concrete, actionable dimension. This paper is a direct response to this ambiguity and argues that transformation must be treated as a distinct strategic domain, one that organizations can systematically work toward. To this end, we review and extend existing paradigms and introduce a new 2x2 framework that moves beyond the oft-cited automation–augmentation duality and offer a matrixed approach that helps leaders visualize and operationalize the full spectrum of human–AI collaboration.

AI has entered organizational theory through three broad paradigms, each reframing the human–technology relationship in distinct ways (Figure 1). The **automation paradigm** positioned AI as a substitute for labor, emphasizing efficiency gains and cost reduction (Brynjolfsson & McAfee, 2014). The **augmentation paradigm** reimagined AI as an amplifier of human capability, elevating productivity while maintaining conventional structures (Davenport & Kirby, 2016). More recently, research has introduced a potential third framing: **collaboration**, in which AI operates as a partner in value creation, enabling organizational forms that move beyond conventional human–machine boundaries (Seeber et al., 2020; Dellermann et al., 2019). While this collaboration perspective reflects the transformative potential of AI, it remains underdeveloped and offers limited guidance for organizational leaders.

**Figure 1**

*Evolution of Human-AI Interaction Paradigms*

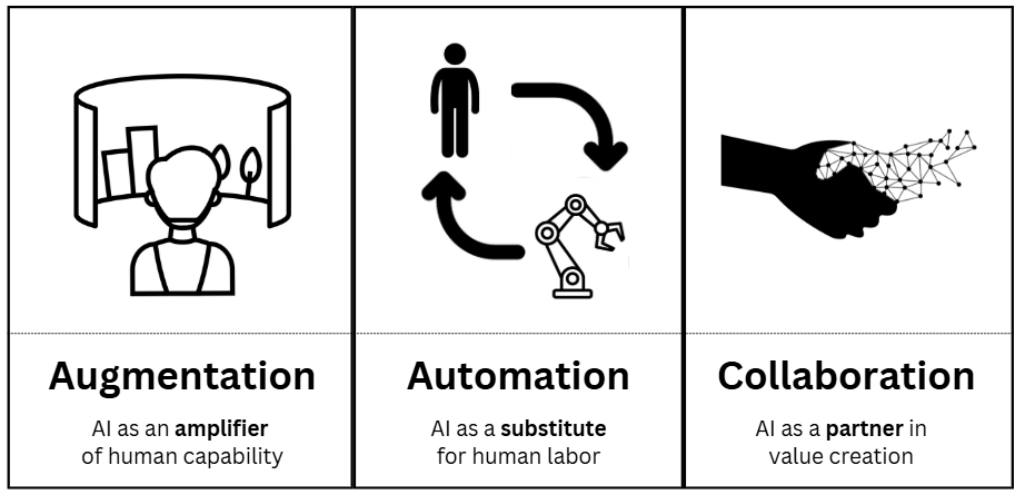

These paradigms together suggest that AI should deliver transformative returns to organizations and individuals, yet results remain modest and uneven. Adoption is now mainstream, with 78% of organizations reporting AI use, but many firms still struggle to convert pilots into scaled, measurable performance gains (McKinsey, 2024). Evidence from MIT indicates that heavy investment has not translated into broad profit impact, and that structural change is concentrated in technology and media while many other sectors remain in experimentation with limited operating model shift (MIT, 2025). Workforce effects are emerging in targeted ways, especially through task redesign and selective reductions in clerical and customer support roles rather than economy-wide displacement (MIT, 2025; McKinsey & Company, 2024). Essentially, there is ongoing adoption without durable value capture, and change without the people and operating systems to absorb it, requiring explicit strategies to translate use into outcomes and to mitigate unintended consequences for workers.

These striking disparities between deployment of AI technologies and performance gains, between adoption and structural change, and between tech-forward optimism and labor displacement—illustrate more than just an execution challenge; it exposes a conceptual gap. Organizations lack the strategic frameworks and operational imagination needed to reconfigure workflows, processes, and priorities in ways that allow AI to deliver on its transformative potential in responsible, ethical ways. Consequently, organizations suffer from what we call **paradigmatic lock-in**, wherein they inadvertently contain transformative AI technologies within outdated paradigms and, in doing so, neutralize its transformative potential and exacerbate labor market volatility through fragmented displacement.

This paper begins to address this conceptual gap that limits how organizations approach AI by fully explicating the collaboration paradigm, not as a third option in a sequence, but as a fundamentally different way of organizing work. Automation and augmentation are often framed as a binary choice; automate a task or enhance it with human input. But as Raisch & Krakowski (2021) argue, they are not opposite; they are interdependent processes that evolve together across time and tasks. We build on this insight even further to propose that transformation itself must be treated as a distinct strategic domain, one that organizations can actively design for. By extending existing perspectives and introducing a matrixed framework, we move beyond the

automation–augmentation dichotomy, which oversimplifies their interdependent nature, and offer a structured approach that helps leaders visualize and operationalize human–AI collaboration in earnest.

## Research Gap: The Collaboration Frontier

Despite decades of scholarship on AI adoption and organizational impact, research has not yet reckoned with the scale of organizational reconfiguration demanded by true human–AI collaboration. What exists today is largely incremental, meaning that AI adoption tends to focus on enhancing productivity within individuals and within systems, such as enhancing individual tasks, automating processes, or substituting labor, without challenging and redesigning the foundational models of how work is organized, coordinated, and valued. This incrementalist approach reveals a few fundamental blind spots:

1. **Instrumental reductionism** – AI is still too often framed as a mere tool, which ignores its growing agency in shaping decisions, workflows, and even strategic direction. This framing limits organizations' ability to understand and effectively leverage AI's emergent capabilities for intended outcomes.

2. **Anthropocentric bias** – Organizational research continues to assume human primacy, overlooking domains where AI already surpasses humans in speed, scale, and analytical precision. As a result, organizations may underutilize AI's strengths or misallocate human labor to asks that are better suited for machine execution.

3. **Static conceptualization** – Integration is treated as a one-off implementation rather than an ongoing cycle of recursive learning and adaptation (Seeber et al., 2020). This view prevents organizations from building systems that evolve with use, leading to missed opportunities for compounding value and increased risk that they will be left behind as technology continues to advance.

These blind spots signal a systemic lag between existing research paradigms regarding human-AI interaction and how leaders are implementing AI in their organizations. Because the latter outpaces the former, organizations lack conceptual scaffolding to redesign jobs, flatten hierarchies, and build teams where AI functions as a true collaborator. Without new frameworks that attend to the aforementioned blind spots, leaders risk locking AI into obsolete command-and-control structures, undermining both employee human and AI capability, and limiting adaptability and long-term impact. More broadly, closing this gap means explicitly preparing for **human–AI collaboration**, where organizational intelligence emerges through networks of people and machines working in optimized harmony. This is the collaborative frontier—an era defined not by incremental adoption, but by reimagined organizational forms where **collective intelligence** is distributed, adaptive, and co-created (Lévy, 1997).

To move beyond these blind spots and meet the demands of the collaborative frontier, organizations need new theory to help inform what is the best way to leverage new AI technologies in a way that leads to that collaborative frontier. While existing human-AI interaction paradigms offer important insights, they are not sufficient for guiding the structural and strategic shifts required for true collaboration. What's needed is a framework that not only accounts for AI's growing role in shaping work, but also helps leaders reimagine how intelligence, agency, and coordination are distributed across people and machines.

This paper addresses three research questions:

- What distinct AI implementation strategies can be observed in contemporary organizations?

- What observable characteristics differentiate these strategies?

- What are the implementation patterns and organizational implications of each strategy?

**Theoretical Framework and Contributions**

This paper integrates theories from information systems, psychology, and labor economics to explain AI's transformation paradox. From information systems, we draw on sociotechnical systems theory (Trist & Bamforth, 1951; Emery & Trist, 1965) to examine how technical AI systems and social work structures must co-evolve for transformation to occur. This perspective reveals why purely technical implementations fail; they optimize the technical subsystem while ignoring the social architecture of work.

From industrial-organizational psychology and organizational behavior, we apply the Job Characteristics Model (Hackman & Oldham, 1976) to analyze how AI affects autonomy, skill variety, and task significance—core determinants of motivation and performance. March's (1991) exploration-exploitation framework explains that organizations default to exploiting AI for incremental gains rather than exploring transformative possibilities. Transactive Memory Systems theory (Wegner, 1987) illuminates how expertise develops through coordinated division of labor between humans and AI. The concept of collective mind (Weick & Roberts, 1993) frames our analysis of organizational intelligence emerging through heedful interrelating between human and artificial agents. From cognitive science, we draw on distributed cognition (Hutchins, 1995) to theorize intelligence as emerging from human-AI interaction rather than residing in either component alone.

From labor economics, we build on task-based models of automation (Autor, Levy & Murnane, 2003) to differentiate routine task substitution from complementary human-AI collaboration. We apply Acemoglu & Restrepo's (2019) framework distinguishing displacement from productivity effects to understand when AI reduces versus amplifies human contribution.

We make three contributions that extend these theoretical foundations:

- A 2x2 strategic framework: We conceptualize AI strategies along two independent dimensions: organizational change (incremental vs. transformational) and human contribution (reducing vs. amplifying). This extends task-based automation models by showing how the same AI capabilities can either substitute for or complement human work depending on organizational design choices.

- The concept of paradigmatic lock-in: Building on March's (1991) exploration-exploitation tension and Leonard-Barton's (1992) core rigidities, we theorize how organizations constrain AI within industrial-era work designs. This violates sociotechnical systems principles of joint optimization, explaining why organizations achieve task-level efficiency without system-level transformation.

- Operationalization of collaborative intelligence: Synthesizing distributed cognition (Hutchins, 1995), Transactive Memory Systems (Wegner, 1987), and collective mind (Weick & Roberts, 1993), we specify three mechanisms through which human-AI collaboration creates value: complementarity, co-evolution, and boundary-setting. This extends both I/O psychology's understanding of team performance and economic models of human-AI collaboration.

**Paper Structure and Approach**

In this theoretical paper we will chart the strategic terrain of AI transformation and identify why most organizations remain trapped in incremental outcomes. First, we will analyze three dominant strategies in current practice—individual augmentation, process automation, and workforce substitution—highlighting their implementation patterns, strengths, and inherent limitations. Second, we will introduce collaborative intelligence as a distinct fourth strategy, define its theoretical mechanisms, and illustrate it through case evidence. Third, we will discuss the implications for organizational design, workforce development, and governance, showing when mixed strategies make sense and when they entrench paradigmatic lock-in. Finally, we will conclude by examining how organizations can prepare for a future defined by transformational, human-amplifying AI rather than efficiency-driven optimization.

**The Present Landscape of AI Strategies: Effectiveness, Context, and Constraints**

**Evidence from Practice**

Building on the theoretical foundations outlined in Section I, this section transitions from the limitations of existing frameworks to how organizations are implementing AI today. We set the stage by outlining three dominant strategies currently in practice, each prevalent yet inherently limited in scope and impact.

- **Individual augmentation**, where AI tools augment specific tasks and reduce individual workload, achieving efficiency gains through gradual reduction of human involvement in routine operations.

- **Process automation**, where AI automates existing workflows and eliminates routine tasks, freeing humans to focus on higher-value activities while maintaining current organizational structures.

- **Workforce substitution**, where AI replaces entire job functions, fundamentally restructuring the organization around automated systems with minimal human oversight.

These strategies yield localized gains in productivity, efficiency, and cost savings, yet fail to achieve broader organizational transformation. Organizations default to incremental improvement of established routines rather than pursuing transformative experimentation (March, 1991), reinforcing existing structures instead of reimagining them (Orlikowski, 1992). This pattern reflects what Christensen (1997) identified as the innovator's dilemma—firms excel at incremental improvements that serve existing paradigms, but struggle with transformational changes that demand new ones. The productivity paradox identified by Brynjolfsson, Rock, & Syverson (2017) highlights this limitation: task-level efficiency improves without translating into systemic performance gains. The financial evidence is stark: despite $30-40 billion in enterprise investment in generative AI, 95% of organizations report no measurable impact on net profits, with only 5% of pilots achieving discernible value (MIT, 2025). The failure stems from overlooking fundamental shifts in work itself: changing skill demands, altered team dynamics, and redistribution of autonomy remain insufficiently theorized and inadequately addressed in practice.

To understand why organizations remain trapped in this pattern and identify pathways forward, we need a framework that maps AI strategies along the dimensions that actually matter for value creation. O'Reilly & Tushman (2013) demonstrate that organizational ambidexterity— simultaneously pursuing exploitation of existing capabilities and exploration of new ones—
requires fundamentally different structures, competencies, and cultures.

Most firms excel at exploitation (incremental refinement) but fail at exploration
(transformational innovation) because they optimize for efficiency, control, and variance
reduction rather than flexibility, autonomy, and experimentation. Figure 2 positions AI strategies
along two critical axes: the degree of organizational change achieved (incremental to
transformational) and the extent to which human contribution is amplified or reduced. This
mapping reveals not only where organizations currently concentrate their efforts—clustered in
exploitative approaches that preserve existing structures—but also the strategic white space
where exploration and human amplification converge.

**Figure 2**

*Framework of AI Transformation Strategies*

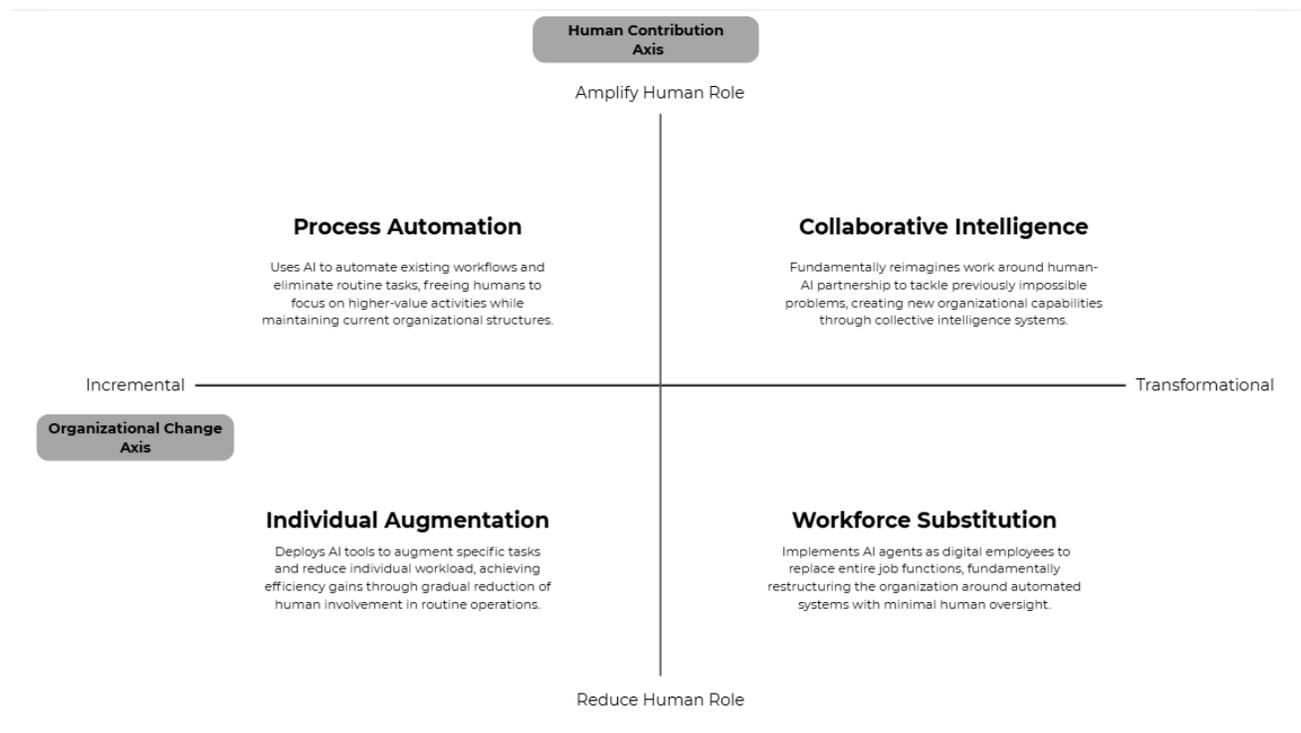

*Note.* Figure 2 situates these approaches on a two-by-two framework. The horizontal axis—the *Organizational Change Axis*—represents a continuum from incremental improvement to transformational change. The vertical axis—the *Human Contribution Axis*—spans from reducing the human role at one end to amplifying human contribution at the other. Together, these axes allow us to position current strategies clearly within the broader landscape of organizational possibilities, showing where they excel and where they fall short of enabling collaborative intelligence.

**The Organizational Change Axis** (horizontal) captures whether AI deployment exploits existing organizational capabilities or explores fundamentally new ones (March, 1991). This dimension ranges from:

- **Incremental**: AI optimizes existing processes, workflows, and capabilities without challenging fundamental organizational assumptions. Observable indicators include: preserved organizational structures, enhanced existing workflows, maintained business models, and optimization of current metrics.

- **Transformational**: AI enables fundamentally new organizational capabilities, business models, or value creation mechanisms. Observable indicators include: restructured organizations, novel workflow designs, new business models, and emergence of previously unmeasurable value streams.

**The Human Contribution Axis** (vertical) captures whether AI implementation reduces or amplifies the human role in value creation. This dimension ranges from:

- **Reduce Human Role**: AI assumes responsibilities previously held by humans, narrowing the scope or significance of human contribution. Observable indicators include:

decreased human decision authority, reduced skill requirements, elimination of human judgment in processes, and performance metrics focused on productivity (i.e. throughput, task completion time, operational costs).

- **Amplify Human Role**: AI enhances human capabilities, expanding the scope or significance of human contribution. Observable indicators include: increased human strategic responsibility, elevated skill requirements, expanded human judgment in complex decisions, and performance metrics focused on human-AI synergy.

From a practical standpoint, these axes are gradients rather than rigid categories. Technologies and the way they are used rarely fall neatly into one quadrant; instead, they often evolve across the framework over time. Consider the trajectory of customer service chatbots. Initial deployments typically augment individual agents by drafting responses or flagging complex queries—incremental improvement that preserves human judgment while improving efficiency. As organizations gain confidence and technology matures, these same systems begin handling entire categories of inquiries autonomously, migrating toward process automation. Eventually, advanced implementations may eliminate entire service tiers, representing workforce substitution.

This migration pattern reveals a critical insight: organizations often drift toward reducing human contribution even when intending to amplify it. The chatbot that began as an augmentation tool becomes an automation system, then potentially a substitution mechanism— each step justified by efficiency gains but collectively eroding human capability and organizational learning. This drift occurs because incremental strategies optimize for immediate performance metrics (response time, cost per interaction) rather than transformational objectives (innovation capacity, adaptive capability). The framework helps organizations recognize this drift and make conscious choices about where they want to position their AI investments rather than defaulting to efficiency-driven migration patterns that may ultimately constrain strategic options.

**Individual Augmentation: The Productivity Play**

**Individual augmentation** refers to the deployment of AI tools that automate discrete tasks, thereby reducing individual workload and generating efficiency gains through the gradual displacement of routine human effort. This strategy has quickly become the most accessible on-ramp to organizational AI adoption, as task-scoped copilots and assistants streamline work without necessitating disruptive structural change. On the 2x2 framework (Figure 2), individual augmentation sits on the incremental end of the organizational change axis, emphasizing individual-level efficiency over organizational-level integration and, in turn, reducing (rather than amplifying) the human role.

Widely adopted applications such as GitHub Copilot, Amazon CodeWhisperer, and Google Smart Compose demonstrate measurable improvements in speed and accuracy, reinforcing the appeal of tools that integrate seamlessly into established workflows (MIT Sloan, 2023; Peng et al., 2023). Market adoption underscores this trend: ChatGPT reached 100 million users in two months, making it the fastest-growing consumer application in history (Reuters, 2023). Microsoft reported that 1 in 3 Fortune 500 companies adopted Copilot for Microsoft 365 within six months of launch (Microsoft, 2024). GitHub Copilot generates approximately 46% of developers' code and has over 1 million paid subscribers (GitHub, 2024). Collectively, these cases illustrate the demand for lightweight, low-friction tools that provide visible productivity gains while avoiding the organizational upheaval associated with deeper transformation.

GitHub Copilot has become a cornerstone example of individual augmentation in action. Survey data reveal that 88% of developers using Copilot report heightened productivity, and performance measures show approximately a 50% reduction in time required to complete boilerplate coding tasks (GitHub, 2023). Experimental studies corroborate these findings: developers explicitly instructed to use Copilot completed assignments 55.8% faster on average; these advantages were especially prominent among junior engineers (Peng et al., 2023; see Figure 3). This evidence demonstrates both the immediate operational value and the democratizing potential of individual augmentation, as less experienced workers are able to perform at levels closer to their senior counterparts. In this sense, GitHub Copilot exemplifies the advantages of the strategy: accelerating execution, reducing cognitive burden, and enhancing accessibility to complex tasks

**Figure 3**

*Time to Task Completion Among GitHub Users (Treated) Vs. Non-GitHub Users (Control)*

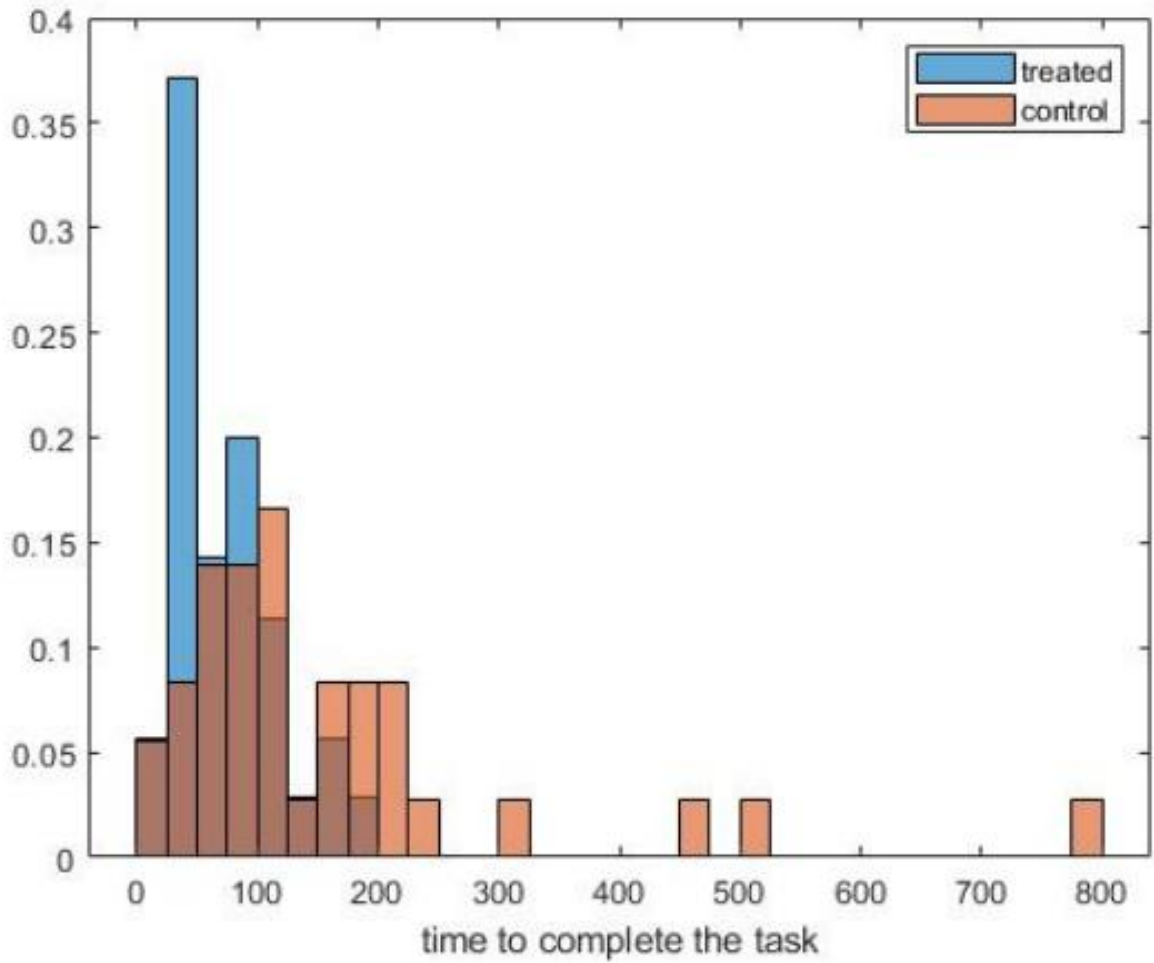

*Note.* Reprinted from "The impact of AI on developer productivity: Evidence from GitHub Copilot," by Peng, S., Kalliamvakou, E., Cihon, P., & Demirer, M., 2023, arXiv:2302.06590. 2302.06590

***Strengths.***

Individual augmentation delivers immediate, measurable productivity gains without requiring organizational restructuring. Brynjolfsson et al. (2023) found that customer service agents using generative AI improved their resolution rates by 14%, with novice workers improving by 35% while experienced workers showed minimal change—demonstrating a democratization effect that narrows skill gaps. Noy & Zhang (2023) documented similar patterns in professional writing tasks: participants using ChatGPT completed tasks 40% faster with higher quality scores, with the weakest performers showing the greatest improvement. Microsoft's Work Trend Index (2023) reported that 70% of Copilot users felt more productive, saving an average of 1.2 hours per week on routine tasks.

The accessibility of individual augmentation distinguishes it from other AI strategies. Organizations can deploy these tools without restructuring workflows, retraining entire teams, or modifying organizational hierarchies, that Rogers (2003) would classify as low complexity and high trialability, factors that accelerate adoption. This rapid deployment capability proves particularly valuable for organizations facing pressure for quick wins or those with limited change management resources.

Employee satisfaction increases when AI handles routine cognitive burden. Noy & Zhang (2023) found that professionals using ChatGPT not only worked faster but reported significantly higher job satisfaction, particularly for previously tedious tasks. The psychological benefits— reduced frustration with routine work, increased time for creative problem-solving, and enhanced sense of competence which aligns with job characteristics theory, which identifies task variety and skill utilization as key drivers of motivation (Hackman & Oldham, 1976). These improvements in both objective performance and subjective experience explain the strategy's rapid adoption across industries.

***Limitations****.*

Individual augmentation's tradeoffs manifest across both individual and organizational dimensions, fundamentally constraining its transformative potential. At the individual level, these tools operate in isolation, supporting task efficiency while failing to build the interdependent capabilities essential for professional development and organizational resilience. Theories of human-AI teaming reveal this gap clearly: Transactive Memory Systems (Wegner, 1987) emphasizes how expertise develops through shared mental models and coordinated division of labor, while collective mind theory (Weick & Roberts, 1993) demonstrates that organizational intelligence emerges through heedful interrelating and coordinated action. Workers operate in parallel silos, each becoming more efficient at discrete tasks but never developing the collaborative workflows or shared understanding that characterize high-performing teams (Salas et al., 2005). This isolation explains why individual augmentation occupies the lower half of the 2x2 framework's vertical axis because the human role is reduced to tool operation rather than amplified through partnership.

The erosion of human agency compounds these limitations. Zuboff's (1988) concept of "informating without empowering" captures this dynamic: technology increases productivity while eroding autonomy. Workers become proficient tool users but lose influence over how tools are deployed, how workflows evolve, and how their roles develop. Contemporary research illustrates this pattern with AI specifically: Lebovitz et al. (2021) found that professionals using AI systems experienced "algorithmic deskilling," where they gradually lost the ability to independently evaluate decisions, defaulting to AI recommendations even when their expertise suggested otherwise. Workers execute AI-suggested actions without understanding the underlying logic or maintaining meaningful discretion, a concept that Pasquale (2015) terms the "black box society," where algorithmic opacity and perceived authority lead users to accept system outputs without meaningful scrutiny or override.

This loss of autonomy can be particularly damaging given the Job Characteristics Model's findings that autonomy drives motivation, performance, and psychological well-being (Hackman

& Oldham, 1976). More concerning still is the deskilling risk: as AI handles increasingly complex tasks, workers may lose fundamental competencies (Lebovitz et al., 2021). Developers relying on code generation tools risk losing understanding of underlying algorithms and system architecture; analysts using AI for data interpretation may forget statistical principles; writers depending on AI assistance may erode their ability to construct arguments from first principles. Jarrahi (2018) argues that workers must develop skills in managing human-AI symbiosis: understanding when to rely on algorithmic recommendations versus human judgment, maintaining awareness of AI limitations, and preserving strategic control over decision-making processes. Without these capabilities, workers become increasingly dependent on tools they neither understand nor control.

Organizational limitations prove equally severe, creating a paradox where tactical efficiency undermines strategic capability. Individual augmentation delivers measurable productivity gains such as faster task completion, fewer errors, and improved accuracy, but these benefits remain trapped at the task level. Because these tools are widely available and easily implemented, their advantages diffuse rapidly across industries, creating what Barney (1991) identifies as competitive parity: all firms achieve similar efficiency improvements, eliminating differentiation and forcing competition on price alone (commoditization). This echoes Carr's (2003) argument that as technologies become ubiquitous, they shift from strategic resources to commoditized necessities. The tools optimize existing workflows without questioning their validity, sitting atop legacy processes and organizational structures that may already be fragmented, inefficient, or misaligned with strategic goals, what Hammer (1990) termed "paving the cow paths," using technology to accelerate fundamentally flawed processes rather than reimagining them. A customer service team using AI chatbots may handle significantly more inquiries, but if the underlying service model remains reactive rather than proactive, the organization achieves efficiency without transformation.

This failure to transform stems from individual augmentation's inability to support organizational learning. Argote and Miron-Spektor (2011) demonstrate that organizational capability emerges through collective knowledge acquisition and transfer, processes that require shared mental models, cross-functional collaboration, and systematic documentation of insights. Individual augmentation tools, designed for solo use, create knowledge silos rather than knowledge networks. These silos prevent organizations from leveraging distributed expertise, creating redundant efforts and missed innovation opportunities (Hansen, 1999). Each worker may become more productive, but the organization fails to capture, synthesize, or scale their learning. The tools fail to populate any of the six organizational memory structures (individuals, culture, transformations, structures, ecology, and external archives), leaving no systematic repository of collective knowledge (Walsh & Ungson, 1991). Without organizational memory, transferable expertise, or collective intelligence, firms experience peripheral blindness: they optimize core operations so intensely that emerging threats and opportunities become invisible (Day & Schoemaker, 2016). Organizations find themselves running faster while standing still, maximizing yesterday's processes while competitors reimagine tomorrow's business models.

While individual augmentation offers an accessible entry point for AI adoption, its inherent limitations—task-level optimization without system transformation, efficiency without strategic differentiation, deskilling without capability building—lead many organizations to pursue more systemic approaches. Process automation emerges as the logical next step, promising to embed AI into workflows rather than individual tasks, though as we will see, this strategy carries its own constraints that may perpetuate rather than resolve the fundamental challenge of achieving transformative rather than incremental change.

## Process Automation: Managed Incrementalism

**Process automation** represents the second major pathway for AI deployment in organizations, focusing on embedding intelligent systems into workflows to optimize, standardize, and streamline operations. Unlike individual augmentation, which typically targets discrete tasks performed by individual workers, process automation operates at the level of end-to-end workflows that involve multiple workers, giving the appearance of being more systemic in scope. In practice, however, these tools often remain siloed within traditional command-and-control structures. Proliferation across users increases coverage but rarely produces the connective tissue of collaborative intelligence or shared organizational memory. On the 2x2 framework (Figure 2), it occupies the incremental side of organizational change axis but leans toward amplifying human roles through retained judgment and oversight. This positions it as a pragmatic choice in risk-sensitive environments, offering operational stability while deferring deeper structural change.

MIT Sloan characterizes this approach as a "small t transformation": a deliberate enhancement of existing operations rather than a reimagining of organizational models (MIT Sloan, 2023; Davenport & Ronanki, 2018). In practice, this means AI automates repetitive, rules-based tasks, while humans remain responsible for interpreting results, handling exceptions, and making decisions. The intent is not to replace human judgment, but to shield it from routine complexity and free up capacity for higher-value work (Trist & Bamforth, 1951; Thompson, 1967).

Organizations implement this through Robotic Process Automation (RPA), intelligent document processing, and automated workflow management systems. Financial services firms using RPA report processing time reductions of 40-70% for routine transactions while maintaining human oversight for complex decisions and regulatory compliance (Davenport & Ronanki, 2018). These systems exemplify what Zammuto et al. (2007) term "affordances": the possibilities for action emerging from technology-organization interactions. The automation handles volume and velocity beyond human capacity while preserving human judgment for context, strategy, and exception management.

***Strengths***.

The strengths of process automation emerge at both the individual and organizational levels. For employees, it reduces cognitive load by offloading repetitive, rules-based tasks such as data entry, scheduling, or document formatting, freeing up mental bandwidth for more judgment-intensive activities like client engagement. Microsoft's Work Trend Index (2023) demonstrates this shift, with 77% of Copilot users reporting reduced 'drudgery' and 1.2 hours weekly saved on administrative tasks that can be redirected to strategic work. While early adoption often involves new overhead such as prompt management and oversight, these findings suggest that workers perceive automation as alleviating routine burden and enabling focus on more judgment-intensive activities.

At the organizational level, automation contributes to greater reliability, continuity, and compliance. These benefits prove especially critical in regulated industries where integrity and standardization of processes are paramount. By embedding AI into standardized workflows, organizations reduce variance in execution and maintain operational consistency across teams and locations. These outcomes reflect how well-designed systems compensate for the limits of human attention and decision-making (March & Simon, 1958), and how automation acts as a buffer against environmental uncertainty by reinforcing predictable, repeatable processes (Thompson, 1967). Contemporary research examines these foundational insights: Shrestha et al. (2021) demonstrate that AI-enabled decision systems specifically address bounded rationality constraints by processing information volumes beyond human cognitive capacity while maintaining consistency across decisions. In this way, process automation not only enhances efficiency but also strengthens organizational resilience.

***Limitations***.

Process automation creates three fundamental constraints that limit its strategic value. First, it generates what Leonard-Barton (1992) terms "core rigidities": the optimization of existing workflows creates path dependencies that actively resist transformation. Enterprise Resource Planning systems exemplify this trap: while delivering operational efficiency, they impose standardized processes that become extremely difficult to modify. Davenport (1998) warned that ERP systems force companies to adapt their processes to the software's embedded logic, while Rettig (2007) documents how these systems create 'concrete', hardened business processes that resist change even when markets shift. The automation crystallizes existing processes into code, including procurement workflows, approval hierarchies, and reporting structures, making deviation exponentially costly.

Second, process automation preserves rather than transcends organizational boundaries. While AI handles routine tasks within departments, it fails to enable the cross-functional intelligence that complex problems demand. The human role narrows to exception handling and oversight, activities that occur within silos rather than across them. This contradicts distributed cognition principles (Hutchins, 1995), which show that organizational intelligence emerges through interaction, not isolation. The preservation of departmental boundaries through automated workflows reinforces existing organizational structures rather than enabling the fluid, networked configurations that complex problem-solving requires.

Third, the redeployment of human talent to strategic work rarely materializes without deliberate intervention. Acemoglu & Restrepo (2019) demonstrate that successful automation requires workforce development investments of 2-3x the technology cost, investments most organizations defer or avoid. Workers lose routine tasks that provided learning opportunities and career pathways without gaining the capabilities for strategic contribution. Organizations must therefore develop what Teece (2007) calls dynamic capabilities, not just in technology deployment but in continuous workforce adaptation, or risk creating a hollowed-out middle tier where neither human judgment nor AI capability effectively operates. While process automation maintains human involvement within existing structures, workforce substitution takes a fundamentally different approach: transforming operations by replacing human roles entirely.

## Workforce Substitution: The Efficiency Imperative and Its Human Cost

**Workforce substitution** represents the third major strategy for organizational AI deployment, characterized by AI systems assuming responsibility for functional domains previously performed by human workers. Unlike individual augmentation or process automation, this strategy does not aim to assist or streamline human work; it seeks to eliminate it. On the 2x2 framework (Figure 2), workforce substitution sits in the bottom-right quadrant, representing a transformational shift in organizational structure while significantly reducing human involvement. It delivers radical change in form, but often without a corresponding evolution in collaborative capability.

This strategy is often driven by economic pressures and competitive dynamics, particularly in commoditized markets where cost reduction is essential for survival (Williamson, 1975). It reflects a broader shift toward minimizing operational overhead by replacing predictable, rules-based human tasks with algorithmic systems. These tasks—often repetitive and easily codified—are especially vulnerable to automation, making substitution an attractive option for organizations seeking scalable efficiency (Arntz et al., 2016). The underlying logic is straightforward: if a task can be consistently defined and executed, it can be delegated to machines at lower cost and higher speed (i.e., routine-biased technological change; Autor, Levy, & Murnane, 2003).

A nuanced case study emerges from manufacturing and logistics, where widespread automation has transformed entire sectors. Manufacturing firms have deployed hundreds of thousands of industrial robots globally, with the International Federation of Robotics reporting over 3.5 million operational units worldwide as of 2022 (IFR, 2023). Amazon exemplifies this transformation at scale, operating over 750,000 mobile robots across its fulfillment network alongside its human workforce (Amazon, 2024). These deployments achieve substantial operational improvements: reduced processing times, lower error rates, and the ability to handle peak demand without proportional workforce expansion. The underlying economic logic follows what Autor, Levy, & Murnane (2003) identified as routine-biased technological change—tasks with clear rules and repetitive patterns become prime candidates for substitution. While companies report efficiency gains and cost reductions, the employment effects vary significantly: some facilities eliminate positions entirely while others shift workers to supervisory and technical roles, though typically at ratios of one new position for every three to ten eliminated (Brynjolfsson & McAfee, 2014).

*Strengths*.

The strengths of workforce substitution are most visible at the organizational level. In highly competitive markets, automating routine work can ensure organizational survival by stabilizing margins and achieving cost structures that human-intensive operations cannot match. Additionally, when implemented thoughtfully, substitution redirects human workers toward activities requiring creativity, complex problem-solving, and emotional intelligence—capabilities that remain difficult to automate. In some cases, substitution strategies generate new roles in adjacent areas, such as AI training, oversight, and exception handling, thus creating alternative employment pathways even as traditional roles are phased out (Acemoglu & Restrepo, 2020).

Beyond economic considerations, substitution eliminates human exposure to dangerous conditions in hazardous environments—chemical processing, mining operations, nuclear facilities—while maintaining operational continuity. The Fukushima Daiichi cleanup deployed PackBot and Warrior robots to handle radioactive debris in areas with radiation levels that would prove fatal to humans within minutes, enabling continuous decontamination work that would otherwise be impossible (TEPCO, 2021). Organizations achieve true 24/7 operations without shift changes, overtime costs, or fatigue-related errors, with automated systems maintaining consistent performance levels that human workers cannot sustain (Brynjolfsson & McAfee, 2014).

From a scalability perspective, substitution enables what economists term "zero marginal cost" expansion—digital services can serve additional customers with minimal incremental investment (Goldfarb & Tucker, 2019). Automated systems handle demand variability without the recruitment, training, and coordination costs associated with human workforce scaling (Brynjolfsson & McAfee, 2014). The net employment effect varies significantly by industry and implementation. Acemoglu & Restrepo (2020) found that industrial robots have a negative effect on employment and wages, with displacement effects outweighing any job creation in technical roles—though the precise ratios vary by sector and region.

***Limitations***.

Despite its economic appeal, workforce substitution carries significant human and societal costs. The impact extends beyond job loss to include community disruption and measurable health consequences, with research linking mass displacement to increased mortality rates and persistent earnings losses (Sullivan & von Wachter, 2009). Estimates of future automation risk vary widely based on methodology: Frey & Osborne (2017) suggest 47% of U.S. jobs face high automation risk when analyzing occupational tasks, while Arntz et al. (2016) find only 9% when accounting for task variation within occupations (Figure 4). Rather than debate precise figures, the critical point remains: millions of workers face displacement, with concentrated impacts on specific communities and demographics. This growing vulnerability has given rise to what Standing (2011) terms "the precariat" which is a new class of workers facing economic insecurity and identity disruption.

**Figure 4**

*Distribution of Automatability in the U.S. (Task-Based vs. Occupation-Based Approach)*

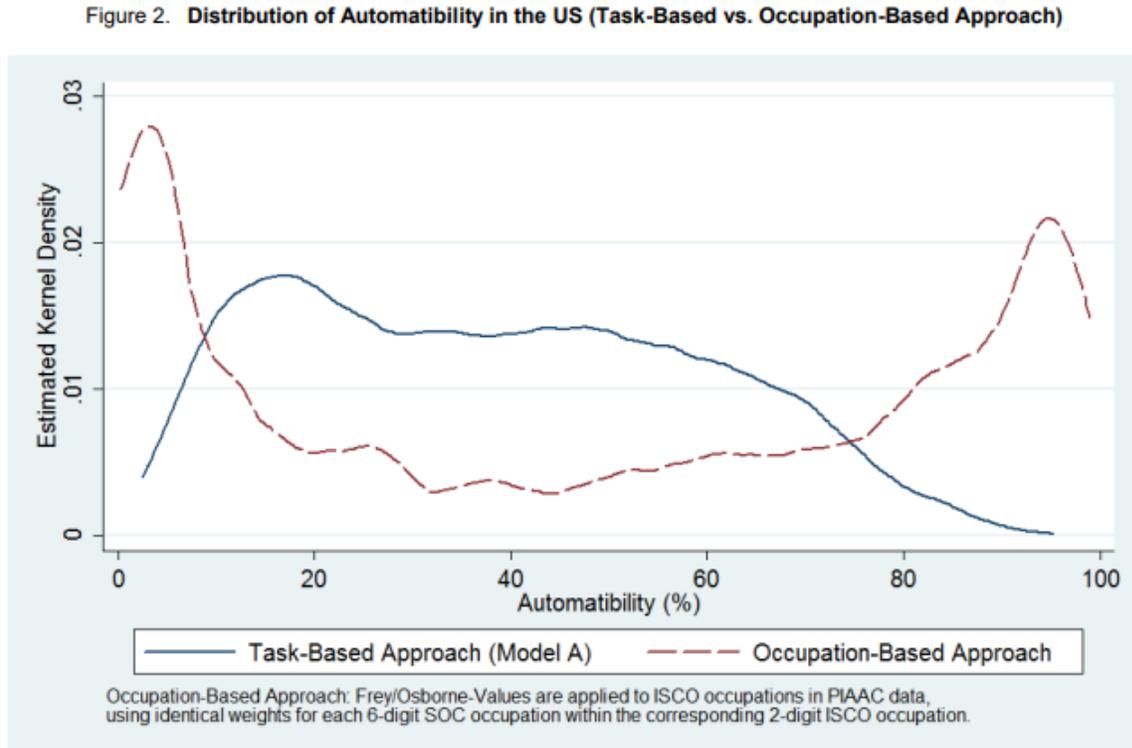

*Note.* Reprinted from "The risk of automation for jobs in OECD countries: A comparative analysis," Arntz, M., Gregory, T., & Zierahn, U., 2016, OECD Social, Employment and Migration Working Papers.

Organizationally, substitution erodes collective mind—the distributed expertise and coordinated action that emerges through human interaction and shared experience (Weick & Roberts, 1993). This collective capability enables organizations to detect subtle anomalies, adapt to unexpected situations, and maintain operational coherence during crises. When organizations eliminate entire job categories, they lose not just individual knowledge but the connective tissue between different forms of expertise—tacit knowledge that cannot be codified or transferred to machines (Polanyi, 1966).

Strategically, the efficiency gains from substitution create a fundamental error: pursuing operational effectiveness without strategic positioning (Porter, 1996). When all competitors adopt similar automation technologies, they converge on identical cost structures and operational processes, eliminating differentiation and forcing competition on price alone. Operational effectiveness, while necessary, proves insufficient for sustained advantage because organizations need unique value propositions that automation alone cannot provide. This convergence toward operational parity through substitution strategies undermines long-term profitability across entire industries, as firms race to automate identical processes rather than developing distinctive capabilities. The result is an industry-wide commoditization trap where efficiency gains benefit customers through lower prices while eroding margins for all competitors equally.

*Workforce Development and Societal Implications.* In workforce substitution contexts, upskilling becomes a critical consideration for organizational sustainability. Organizations pursuing this strategy face both moral and practical questions about supporting workforce transitions. IBM's SkillsBuild program, which aims to educate 30 million people by 2030, illustrates one approach to addressing displacement at scale (IBM, 2020). Research by Illanes et al. (2018) suggests that successful transition programs typically require 18-24 months of support, combining technical training, career counseling, and financial assistance. These findings indicate that workforce substitution may be better understood not as a one-time cost reduction but as an ongoing commitment requiring continuous investment in human capital development.

Workforce substitution represents the furthest departure from collaboration, eliminating rather than augmenting human contribution. Yet even within this model, opportunities for human-AI partnership can emerge in roles such as oversight, training, and exception handling. While these new positions develop alongside automation, they often require different skill sets and tend to be filled through external labor markets, resulting in incumbent workers being displaced rather than reintegrated. The societal implications raise questions about appropriate policy responses, with some advocating for universal basic income, portable benefits, and education system reform. The Danish "flexicurity" model, combining labor market flexibility with strong social protections, offers one potential framework for managing technological transitions (Madsen, 2004). These challenges suggest that organizations might benefit from considering their broader role in shaping workforce transitions rather than viewing displacement as solely a governmental concern.

## Hybrid Configurations: Balancing Multiple AI Strategies

These three strategies—individual augmentation, process automation, and workforce substitution—constitute the practical toolkit of AI transformation today and, importantly, they work. Moreover, organizations can achieve value not by choosing one approach but by strategically combining them based on functional needs, competitive pressures, and workforce capabilities. JPMorgan Chase exemplifies this mixed-strategy success: deploying individual augmentation through code-completion tools for developers, process automation via COiN for contract analysis (saving 360,000 hours annually), and selective workforce substitution in routine transaction processing—all while investing $12 billion annually in technology. This portfolio approach demonstrates how thoughtful strategy mixing can deliver both efficiency gains and workforce development.

The prevalence of such hybrid approaches reflects organizational wisdom born from practical constraints; these three strategies offer proven paths to value creation while managing risk. Most organizations will inhabit this mixed-strategy space for the foreseeable future as they navigate technical debt, regulatory requirements, workforce transitions, and cultural change. Historical research on technology adoption cycles (Rogers, 2003) suggests that organizational transitions often take years to stabilize, yet recent evidence indicates that AI is compressing traditional adoption timelines, with consumer-led demand and rapid diffusion accelerating enterprise uptake (McKinsey, 2023). This is not strategic failure but pragmatic adaptation in action—organizations learning to reconfigure resources while maintaining operational stability (Eisenhardt & Martin, 2000).

Yet each strategy carries inherent limitations that compound over time. Individual augmentation plateaus at task-level productivity without system-level transformation. Process automation preserves organizational silos even as it optimizes workflows within them. Workforce substitution achieves cost reduction but erodes the very human capital needed for innovation and adaptation. These are not fatal flaws but bounded solutions—necessary for today's competition but insufficient for tomorrow's challenges. The very capabilities that drive an organization's current success can become barriers to future transformation, as they are often optimized for existing conditions rather than emerging ones (Christensen & Overdorf, 2000).

The critical question is not how to perfect these strategies but what lies beyond them. **The hybrid-strategy approach, while stable, represents what complexity theorists call a "local maximum"—the best solution within current constraints but not the best solution possible** (Kauffman, 1993). Organizations optimizing within this paradigm may achieve operational excellence, but they risk reinforcing the very frameworks that limit transformative potential. This is the essence of paradigmatic lock-in: by continuing to refine strategies rooted in legacy assumptions and frameworks, firms inadvertently constrain their ability to reimagine what AI can enable. The larger transformation opportunity emerges only when we fundamentally reconceptualize the relationship between human and artificial intelligence.

This brings us to the fourth quadrant of our framework—collaborative intelligence— which represents not the next incremental step but a fundamentally different organizing logic. The next section explores this frontier—not as distant speculation but as an emerging reality being pioneered by organizations that recognize that sustainable competitive advantage no longer comes from optimizing the human-machine divide but from dissolving it entirely.

## Collaborative Intelligence: An Emerging Strategic Frontier

On the 2x2 framework, collaborative intelligence is positioned in the quadrant defined by transformational change and amplification of the human role. Unlike the other strategies, which automate or constrain work within existing structures, collaborative intelligence expands the frontier by enabling new forms of human–AI teaming. Its placement on this framework reflects its theoretical positioning as addressing different organizational objectives than the other strategies.

### Defining Collaborative Intelligence

**Collaborative intelligence** can be defined as an organizational pattern where humans and AI systems function as interactive partners in generating solutions, making decisions, and creating value. Rather than a linear extension of automation, it is a deliberately hybrid model that combines complementary human and machine strengths—human social reasoning, judgment, and creativity with algorithmic scale, speed, and pattern detection (Dellermann et al., 2019; Kamar, 2016; Wilson and Daugherty, 2018). Conceptually, it builds on collective-intelligence research showing that groups exhibit a reliable collective intelligence factor, or c-factor, which functions as a group-level analogue to individual IQ and predicts performance across a wide range of tasks (Levy 1997; Woolley et al., 2010) and that social-cognitive capabilities (e.g., theory of mind) help explain why some teams consistently outperform others both online and face-to-face (Engel et al., 2014). Extending these insights to socio-technical teams, contemporary work argues for designing roles, norms, and interfaces so that AI systems integrate into team workflows as collaborators rather than mere tools (Seeber et al., 2020). Three empirically observable patterns distinguish collaborative intelligence from other AI strategies:

*Complementarity*, derived from human-computer interaction theory (Hollan et al., 2000) and task-based models of automation (Autor et al., 2003), manifests as dynamic task allocation based on comparative advantage rather than automation feasibility. Unlike augmentation (where AI assists predefined human tasks) or substitution (where AI replaces human functions), complementarity involves continuous reconfiguration of roles based on situational demands. AI systems handle combinatorial exploration, pattern recognition across massive datasets, and rapid hypothesis generation, while humans provide contextual interpretation, ethical judgment, and creative problem framing. This division transcends simple task splitting; it operationalizes what Licklider (1960) envisioned as "man-computer symbiosis" where intelligence emerges from interaction rather than residing in either component.

*Boundary-setting*, grounded in algorithmic governance literature (Danaher et al., 2017) and human-in-the-loop systems research (Mosqueira-Rey et al., 2023), preserves human agency over strategic parameters while delegating tactical execution to AI. Humans retain authority over value definitions, ethical constraints, risk thresholds, and strategic objectives. AI operates autonomously within these boundaries, making tactical decisions and executing operations at speeds and scales beyond human capacity. This preserved strategic control distinguishes collaboration from substitution strategies where human judgment is eliminated entirely, reflecting what Simon (1960) described as the separation between programmed and non-programmed decisions.

*Co-evolution*, synthesized from organizational learning theory (March, 1991) and human-AI teaming research (Seeber et al., 2020), describes the bidirectional learning dynamic between human and AI agents. Humans develop new competencies in query formulation, output interpretation, and strategic AI deployment, while AI systems improve through human feedback, correction, and contextual guidance. This mutual adaptation differentiates collaborative intelligence from static automation, where capabilities remain fixed post-deployment. The co-evolutionary process creates what Teece (2007) calls dynamic capabilities: the capacity to continuously reconfigure competencies as conditions change. As our empirical analysis will demonstrate, this component remains largely theoretical in current implementations.

Current implementations cluster in domains, requiring synthesis of computational power with human judgment. In pharmaceutical R&D, organizations are restructuring discovery processes around human-AI partnership rather than sequential handoffs. Insilico Medicine's advancement of a novel drug candidate from target identification to preclinical validation in 18 months demonstrates this pattern: AI systems explored molecular space and predicted interactions while scientists provided biological context, clinical design expertise, and regulatory navigation (Zhavoronkov et al., 2019). BenevolentAI's collaboration with AstraZeneca follows similar principles, with AI generating hypotheses for drug repurposing while humans determine clinical trial parameters and ethical boundaries (AstraZeneca, 2020). These pharmaceutical examples reveal a critical distinction: collaborative intelligence requires organizational restructuring, not merely tool deployment. Traditional drug discovery follows linear stages:

target identification, lead optimization, preclinical testing, with clear handoffs between specialized teams. The collaborative model dissolves these boundaries, creating integrated human-AI teams that iterate rapidly across all stages simultaneously. Scientists no longer simply review AI outputs; they engage in continuous dialogue with algorithms, refining parameters, challenging assumptions, and redirecting exploration based on emerging insights.

*Strengths.*

Collaborative intelligence theoretically addresses limitations inherent in other strategies. By preserving and amplifying human expertise rather than replacing it, organizations maintain the adaptive capacity and contextual judgment that pure automation sacrifices. The approach generates what Barney (1991) terms sustained competitive advantage through valuable, rare, inimitable, and non-substitutable capabilities making the unique configuration of human-AI partnership difficult for competitors to replicate. Research on hybrid intelligence systems suggests enhanced problem-solving when humans and AI work interactively rather than separately (Dellermann et al., 2019), though systematic organizational evidence remains limited.

*Limitations.*

The gap between theoretical promise and observed practice reveals substantial implementation challenges. Achieving collaborative intelligence requires more than technological deployment; it demands fundamental organizational transformation. Hierarchical structures must evolve toward networked configurations where expertise flows to problems regardless of formal position (Burns & Stalker, 1961). Cultural shifts prove equally demanding; employees must reconceptualize AI from threatening automation to collaborative partner, requiring new models of trust, accountability, and performance evaluation (Glikson & Woolley, 2020).

Critical uncertainties persist as no validated metrics distinguish genuine collaborative intelligence from sophisticated automation. Systematic documentation of successful implementations remains absent beyond vendor claims and isolated cases. The specific capabilities, resources, and timelines required for transformation lack empirical specification. Whether collaborative intelligence represents an achievable organizational strategy or remains a theoretical ideal cannot be determined from current evidence. These limitations position collaborative intelligence as an emerging possibility rather than an established practice, a frontier that organizations are beginning to explore but have not yet fully mapped. To further understand the frontier that organizations are beginning to explore but have not yet fully mapped. To examine whether these theoretical components manifest in practice and identify the specific barriers to their implementation, we now turn to empirical analysis of documented AI deployments across multiple sectors.

**Collaborative Intelligence Case Studies**

To systematically examine collaborative intelligence implementations, we apply the framework developed in Section 3.1 to documented AI deployments across multiple sectors. This framework—comprising complementarity, co-evolution, and boundary-setting—serves as an analytical lens to identify which theoretical components organizations successfully operationalize versus which remain unrealized. Our analysis draws exclusively from peer-reviewed publications, government reports, and SEC filings to ensure empirical rigor. This methodological approach reveals a consistent pattern: mature implementation of components that preserve existing organizational structures alongside systematic absence of those requiring fundamental transformation.

***Illustrations of Complementarity.***

Organizations demonstrate mature implementation of complementarity, establishing clear divisions between human strategic judgment and AI computational processing that align with current technical capabilities.

In pharmaceutical research, Atomwise's Artificial Intelligence Molecular Screen (AIMS) exemplifies operational complementarity. Published results in Nature Scientific Reports (2024) document the system screening 15 quadrillion compounds across 318 protein targets, identifying "bioactive scaffolds across a wide range of proteins, even without known binders, X-ray structures, or manual cherry-picking of compounds." The division of labor follows comparative advantage: AI performs combinatorial exploration at scales impossible for humans ($>10^{16}$ molecular combinations), while human researchers provide target selection, biological validation, and strategic research direction. This represents complementarity as theorized: dynamic task allocation based on distinct capabilities rather than simple automation.

Healthcare applications demonstrate similar patterns. IDx-DR, the first FDA-cleared autonomous diagnostic AI, achieved 87.2% sensitivity and 90.7% specificity for diabetic retinopathy detection across 900 subjects (Abràmoff et al., 2018). The system operationalizes complementarity through explicit role division: primary care staff operate imaging equipment following standardized protocols, AI performs diagnostic analysis without specialist involvement, and ophthalmologists focus exclusively on treatment-requiring cases. This restructuring expanded screening access to primary care settings previously lacking specialist availability, demonstrating how complementarity can create new capabilities rather than only optimizing existing ones.

Financial services exhibit complementarity at scale. BlackRock's Aladdin platform, processing $21.6 trillion in assets, demonstrates systematic task division documented in the Harvard Business School case analysis (Lerner & Tufano, 2018). During COVID-19 market volatility, Aladdin computed risk scenarios across millions of securities simultaneously, a computational feat impossible for human analysts, while portfolio managers applied contextual understanding of regulatory changes, client psychology, and market sentiment that AI cannot replicate. The system performs calculations in minutes that would require months of human analysis, while humans provide intuitive pattern recognition and relationship management capabilities.

### Illustrations of Boundary-Setting.

Boundary-setting appears comprehensively implemented in regulated industries where compliance frameworks mandate human oversight and accountability mechanisms.

Other examples from financial services demonstrate the most developed boundary-setting frameworks. The Federal Reserve's Model Risk Management guidance (SR Letter 11-7, 2011) requires banks to "document model limitations and assumptions" and "establish clear escalation procedures for model overrides." Implementation evidence from the Richmond Fed (2023) reveals explicit authority structures: AI recommendations require human approval before execution, human analysts possess unilateral override power regardless of AI confidence levels, and must document justifications when deviating from AI suggestions. This creates explicit boundaries: AI generates recommendations but cannot execute trades, while humans retain sole execution authority and legal accountability.

Healthcare systems exhibit equally robust boundary-setting. Viz.ai's stroke detection platform, analyzed across 14,116 alerts at 166 facilities (Hassan et al., 2024), demonstrates operationalized control mechanisms. While AI automatically alerts neurologists to suspected large vessel occlusions within 4 minutes, the system cannot initiate treatment protocols or modify patient records. Physicians possess three levels of control: they can disable the AI system entirely, adjust sensitivity thresholds between 60-95%, or exclude specific imaging protocols from AI analysis. The AI has zero autonomous decision rights; it functions purely as an alerting system under complete physician control.

Government applications reveal boundary-setting in high-stakes contexts. Palantir's deployment documentation (GAO Report B-412746, 2016) describes systems that "combine all intelligence software/hardware capabilities" with mandatory human analytical oversight. Project Maven explicitly prohibits autonomous targeting: "AI will not be selecting a target [in combat]...What AI will do is complement the human operator" (DoD, 2017). These boundaries reflect both ethical considerations and recognition of AI limitations in contexts requiring contextual judgment.

### *Illustrations of Co-Evolution.*

Co-evolution is recursive learning where human expertise enhances AI capabilities while AI insights transform human understanding. This domain remains largely unimplemented across examined sectors. This absence appears to reflect both technical constraints and organizational barriers.

Limited evidence exists primarily in research environments. Autodesk's "Hybrid Intelligence" study (ASME Journal of Mechanical Design, Song et al., 2023) documents "a human-centered continual training loop to seamlessly integrate AI-training into the expert's task workflow." Eight architects using the system showed measurable behavioral adaptation: designers' confidence adjusted asymmetrically to AI performance, decreasing with poor results but not increasing proportionally with good results (Zhang et al., 2023). However, these remain controlled experiments rather than production implementations, suggesting technical or organizational barriers to scaling co-evolutionary mechanisms.

Pharmaceutical and financial sectors, despite sophisticated AI deployments, show minimal co-evolution evidence. BenevolentAI's published work (Frontiers in Pharmacology, Stebbing et al., 2021) mentions "human-machine interaction which enabled scientists to identify relevant biological contexts" but provides no metrics for systematic learning transfer. BlackRock's Aladdin documentation lacks mechanisms for portfolio manager insights to modify algorithmic behavior or for AI patterns to fundamentally restructure trading expertise. This absence is particularly notable given these sectors' technical sophistication and resources.

***Synthesis of Implementation Patterns.***

Table 1 synthesizes our analysis across nine major AI implementations, revealing a striking pattern: while complementarity appears in every documented case and boundary-setting emerges wherever regulation requires it, co-evolution remains absent from production deployments.

**Table 1.**

*Implementation Analysis of Case Studies for Collaborative Intelligence Components by Industry*

| | The Three Components of Collaborative Intelligence | | |
|---|---|---|---|
| **Case Study** | **1. Complementarity** | **2. Boundary-Setting** | **3. Co-Evolution** |
| **Pharmaceutical Sector** | | | |
| Atomwise AIMS | ✓ Present | Not documented | Not documented |
| BenevolentAI | ✓ Present | Not documented | Not documented |
| Insilico Medicine | ✓ Present | Not documented | Not documented |
| **Healthcare Sector** | | | |
| IDx-DR | ✓ Present | ✓ Present | Not documented |
| Viz.ai | ✓ Present | ✓ Present | Not documented |
| **Financial Sector** | | | |
| BlackRock Aladdin | ✓ Present | ◑ Partial | Not documented |
| Federal Reserve | ✓ Present | ✓ Present | Not documented |
| **Other Sectors** | | | |
| Palantir Foundry (Government) | ✓ Present | ✓ Present | Not documented |
| Autodesk Platform (Engineering) | ✓ Present | Not documented | ◑ Partial |

*Note.* ✓ Present = comprehensive documented evidence in peer-reviewed or official sources; ◑ Partial = limited or incomplete evidence; Not documented = absence of evidence in examined sources. Data sources include Nature Scientific Reports (2024), FDA 510(k) clearance data, Harvard Business School Case 9-217-032, Frontiers in Stroke (2024), GAO-16-504SP, ASME Journal of Mechanical Design (2023), SR Letter 11-7, Frontiers in Pharmacology (2021), and Nature Biotechnology (2022).

Our cross-industry analysis reveals that collaborative intelligence, as theoretically conceived, remains fragmented in practice. The systematic presence of complementarity and selective implementation of boundary-setting, contrasted with the paucity of co-evolution across Table 2, provides empirical evidence for our paradigmatic lock-in thesis. Organizations successfully implement AI components that preserve existing structures: task division that maintains role hierarchies, governance that preserves decision rights. While failing to develop mechanisms that would enable genuine human-AI partnership and organizational transformation.

This component gap reveals the mechanism through which organizations achieve tactical AI success while missing transformational potential. Whether this pattern reflects temporary technical constraints that future advances will resolve, or fundamental organizational limitations that will persist regardless of technological progress, remains an open empirical question. The evidence suggests, however, that achieving collaborative intelligence requires not merely technological advancement but organizational willingness to reimagine how intelligence, expertise, and learning flow through human-machine networks; a transformation that current implementations have yet to achieve.

## Discussion and Implications

The transformation landscape of AI implementation reveals a striking pattern: organizations deploy transformative technology yet achieve predominantly incremental outcomes. Our analysis of contemporary strategies (individual augmentation, process automation, workforce substitution, and the emerging collaborative intelligence model) exposes a fundamental constraint we term paradigmatic lock-in, wherein organizations unconsciously contain AI within industrial-era work designs that neutralize its transformative potential. Table 2 synthesizes these strategic patterns, highlighting how each approach generates value while simultaneously reinforcing the organizational structures that limit transformation. The remainder of this section examines the practical implications of these patterns for organizational strategy, identifies the boundary conditions that determine when different approaches become viable, explores the limitations of our framework, and proposes directions for future research that might help organizations transcend the constraints that currently bind them.

**Table 2**

*AI Implementation Strategies and the Paradigmatic Lock-in Problem*

| Strategy | Description | Strengths | Tradeoffs | Paradigmatic Lock-in |
|---|---|---|---|---|
| **Individual Augmentation** | Deploy AI tools to automate discrete tasks and reduce individual workload | Immediate productivity gains; rapid deployment without restructuring; democratizes expertise (novices approach expert performance) | Algorithmic dependence erodes human judgment; competitive parity eliminates advantage; knowledge silos prevent organizational learning | Trapped in task-based thinking: assumes current tasks are correct and just need acceleration, preventing reimagination of work itself |
| **Process Automation** | Embed AI into workflows to optimize and standardize operations | Reduces variance and ensures compliance; frees mental bandwidth for judgment-intensive work; strengthens operational resilience | Creates codified systems resistant to change; preserves silos; strategic redeployment rarely materializes | Cements status quo processes, leading to a rigid structure where innovation and adaptation are constrained |
| **Workforce Substitution** | Replace human workers with AI systems | Radical cost reduction; enables 24/7 operations; eliminates human exposure to dangerous conditions | Loses tacit knowledge and learning capacity; drives industry homogenization eliminating differentiation; creates workforce precarity | Achieves structural change by eliminating human contribution, sacrificing collaborative potential for efficiency |
| **Collaborative Intelligence** | Humans and AI function as interactive partners in value creation | Theoretically generates novel capabilities and sustains competitive advantage; preserves and amplifies human expertise | Requires fundamental restructuring beyond technology; demands 2-3x investment in organizational change; lacks empirical validation of complete implementation | Escapes lock-in conceptually but remains unachieved in practice |

*Note.* Paradigmatic lock-in occurs when organizations contain transformative AI within industrial-era work designs. The first three strategies achieve their strengths by accepting this constraint; collaborative intelligence requires transcending it.

**Practical Implications**

Organizations should approach AI deployment as a managed portfolio, explicitly positioning each initiative within the transformation-by-human-contribution framework and establishing strategy-specific evaluation criteria. Prior to implementation, leaders should document an initiative's intended quadrant and define observable indicators for movement along each axis: shifts in decision authority, modifications to reporting structures, emergence of new revenue streams, or changes in skill requirements. This preregistration prevents post-hoc rationalization of outcomes and ensures accountability to stated objectives. For initiatives positioned as incremental (individual augmentation or process automation), success metrics should emphasize operational indicators: variance reduction, compliance rates, processing time, and cost per transaction rather than conflating local productivity gains with organizational transformation. Conversely, initiatives claiming transformational impact must demonstrate structural evidence: reconfigured workflows that cross previous boundaries, altered coordination mechanisms between departments, new governance models that redistribute decision rights, or documented emergence of previously impossible capabilities. Resource allocation should reflect these distinctions. Transformational initiatives require budgets encompassing not only technology infrastructure but also job redesign, workforce retraining, data ontology development, and incentive realignment (Acemoglu & Restrepo, 2019).

Given the absence of empirical evidence for fully realized collaborative intelligence, organizations should treat this fourth quadrant as a potential future state rather than an immediately achievable target. However, current AI decisions can be made with this horizon in mind. When evaluating AI initiatives, leaders should assess whether implementations preserve future optionality for human-AI partnership or constrain it through excessive automation. This means maintaining human expertise even as AI assumes routine tasks, documenting decision rationales that could eventually train AI systems, and selecting platforms that expose their reasoning processes rather than operating as black boxes. Organizations can build toward collaborative intelligence systematically by establishing pilot programs where humans and AI jointly tackle complex problems, creating feedback mechanisms that capture how human judgment improves AI outputs, and developing metrics that value collaborative outcomes over individual productivity. Most critically, organizations should resist the temptation to mislabel managed automation as transformational collaboration, recognizing that efficiency gains alone do not constitute partnership. Most critically, organizations should resist the temptation to frame every AI deployment as transformational collaboration when it represents managed automation. The empirical record suggests that for most organizations, well-executed incremental strategies with clear efficiency objectives remain more defensible than premature claims of human-AI partnership. Building technical infrastructure, organizational capabilities, and cultural readiness for collaborative intelligence will likely require years of deliberate investment. Until then, honest assessment of current limitations coupled with strategic preparation for future possibilities represents the most prudent path forward.

**Limitations**

 This framework provides a descriptive taxonomy rather than a predictive model. While it maps current AI deployment strategies and identifies their characteristic patterns, it cannot specify which organizations should pursue which quadrant or predict success rates for different approaches. The two axes deliberately simplify complex organizational phenomena into binary dimensions; in practice, transformation exists on a continuum and human contribution varies by task, department, and implementation phase. Organizations may simultaneously occupy multiple quadrants across different functions, and initiatives may migrate between quadrants over time in ways our static framework cannot fully capture. The concept of paradigmatic lock-in itself faces important limitations. First, it may overstate organizational agency by implying that firms consciously choose to remain within industrial-era paradigms when market forces, regulatory requirements, or technical constraints may dictate these boundaries Second, what we interpret as lock-in may, in some cases, represent rational optimization within current technological and economic realities rather than cognitive limitation. This is especially evident in high-risk industries such as healthcare, financial services, and defense, where cautious progress is often the wiser choice. A more balanced approach is to compartmentalize high-risk functions while creating safe arenas such as sandboxes or external labs where organizations can experiment with collaborative intelligence, simulate scenarios, and red-team emerging applications. Such protective strategies enable pattern recognition and collective learning without compromising core systems. Third, the paradigmatic lock-in framing assumes that transformation is inherently desirable, when evidence suggests that many organizations achieve sustainable success through incremental improvement. The concept risks pathologizing pragmatic decision-making as strategic myopia.

The empirical evidence for collaborative intelligence remains limited to isolated case studies and vendor reports rather than systematic organizational research. Our analysis identifies complementarity and boundary-setting in production deployments but finds co-evolution largely absent, raising questions about whether this component is technically premature, organizationally infeasible, or simply unmeasured by current evaluation methods. The pharmaceutical and financial services examples, while instructive, may not generalize to sectors with different regulatory environments, competitive dynamics, or technical infrastructures. Manufacturing organizations facing commodity competition may find workforce substitution economically necessary regardless of its limitations, while professional services firms may achieve sustainable advantage through individual augmentation alone. Our analysis relies on publicly available documentation, which may underrepresent failed implementations and overemphasize successful deployments. Organizations rarely publish detailed accounts of AI initiatives that failed to achieve intended outcomes, creating potential survivorship bias in our case selection. The co-evolution mechanisms we propose (learning reviews, parameter modification, capability emergence) lack validated measurement instruments, making it difficult to distinguish genuine human-AI partnership from sophisticated automation with human oversight. Additionally, the framework does not address critical implementation variables such as data quality, technical debt, vendor lock-in, or the availability of AI talent, all of which may determine outcomes more strongly than strategic positioning. These constraints suggest that while the framework offers useful organizing principles for AI strategy, organizations should treat it as one input among many rather than a comprehensive guide to implementation decisions.

**Future Research Directions**

The most critical research priority is establishing whether genuine cases of collaborative intelligence are emerging in practice or whether this fourth quadrant remains a theoretical ideal. Our analysis found limited evidence beyond vendor claims and isolated examples, raising questions about whether technical, organizational, or measurement barriers prevent its realization. The mechanisms of paradigmatic lock-in require systematic investigation: what interventions successfully break path dependencies in AI adoption, how organizations unlearn industrial-era assumptions about work division, and whether external shocks or competitive pressures can catalyze genuine transformation. Longitudinal studies should track whether organizations follow predictable trajectories between quadrants or remain locked in initial positions, and whether mixed strategies prove more viable than pure positioning. Industry-specific research should examine whether strategic patterns vary systematically across sectors or whether common constraints affect all organizations regardless of context. Finally, distributional effects demand attention beyond efficiency metrics: which roles benefit versus suffer within organizations, which sectors capture versus lose value, and how AI deployment reshapes employment patterns and economic structure. These investigations will determine whether organizations can transcend current incremental strategies or must accept that human-AI interaction will remain bounded.

## Conclusion

Organizations implementing AI face choices that raise fundamental questions about the nature of work and value creation.  The evidence reveals a troubling pattern: despite revolutionary potential, AI deployment follows paths of least resistance (i.e., automating familiar tasks, augmenting individual productivity, substituting routine labor), without questioning whether the underlying systems of work should be preserved at all. These choices carry profound implications for workforce development, economic opportunity, and human agency in the workplace. The framework presented here suggests that paradigmatic lock-in constrains not just organizational performance but our collective imagination about what work could become in an AI-mediated economy. The urgent question is not how to optimize existing work structures with AI, but rather whether those structures serve human flourishing in an era where intelligence can be genuinely distributed across human-machine networks.

The current strategic landscape reveals organizations defaulting to AI strategies that reinforce rather than reimagine industrial-era work patterns. Individual augmentation accelerates task completion while potentially reducing workers to efficient tool operators, but what are the long-term effects on the autonomy that drives motivation and innovation? Process automation preserves hierarchical decision-making even as it eliminates routine work that traditionally provided entry-level employment, how will this affect skill development pathways? Workforce substitution reduces costs by eliminating human judgment from areas where it may be most critical; what essential capabilities might organizations lose in pursuit of efficiency? Each strategy optimizes metrics of efficiency, yet questions remain about whether this erodes the human capabilities—creativity, judgment, empathy, contextual reasoning—that remain irreplaceable and provide meaning in work.

Collaborative intelligence offers an alternative vision where AI might amplify rather than diminish human contribution, though questions about its achievability persist. This transformation demands substantial investment not just in technology but in developing new skills, redesigning workflows, and reimagining career progressions. The observed 2-3x cost multiplier for organizational change versus technology raises important questions: what determines whether such investments yield returns? How do organizations develop the meta-capabilities needed for human-AI collaboration? What new forms of organizing might emerge?

The implications extend beyond individual organizations to societal questions about employment, education, and economic equity. If collaborative intelligence becomes the dominant paradigm, educational systems must evolve from teaching specific skills to developing meta-cognitive capabilities: the ability to work with, through, and alongside AI systems that will themselves continuously evolve. Workforce development programs must shift from retraining displaced workers for new tasks to preparing them for fundamentally different ways of creating value. Labor policies designed for clear human-machine boundaries become obsolete when intelligence is distributed across hybrid networks where contribution and accountability blur. These are not distant challenges; they are immediate choices being made in every AI implementation decision, each reinforcing or reshaping assumptions about human value in economic systems.

This paper began with a paradox: why has AI failed to deliver transformative returns despite revolutionary potential? Our framework reveals that transformation involves organizational alongside technical challenges, raising questions not just about how we work but why we preserve certain structures and what values they encode. Organizations experimenting beyond paradigmatic lock-in may pioneer new forms of human-machine collaboration that could define the next era of work. Yet questions remain: what determines which organizations successfully navigate this transition? What are the consequences of continued optimization within inherited constraints versus the risks of pursuing transformation?

The strategic patterns documented here suggest organizations face not a choice between right and wrong strategies, but between different trade-offs whose long-term implications remain uncertain. As AI capabilities advance and organizational experience accumulates, how these patterns evolve will shape fundamental aspects of economic organization and human work. Will collaborative intelligence transition from theoretical possibility to practical reality? Will hybrid strategies remain dominant? Will entirely new patterns emerge? The choices organizations make today about AI strategy are fundamentally choices about human agency, dignity, and potential in tomorrow's economy—decisions too consequential to be left to technological determinism or strategic drift.